# HTS Dynamo Flux Pump: The Impact of a Ferromagnetic Stator Substrate

Vladimir Sokolovsky and Leonid Prigozhin

*Abstract*— **HTS dynamo magnetic flux pumps are perspective devices for contactless charging the superconductor magnets and coils. In this work we investigate the influence of a ferromagnetic substrate of a coated conductor used as the pump stator. We use the thin shell model of a coated conductor with a ferromagnetic substrate and show that such a conductor increases the pump generated voltage if the superconducting layer is between the rotor and the substrate. Chebyshev spectral method is employed for numerical solution. Using simulation results for problems with a given transport current we also derive a simple analytical description for feeding a current to a coil.**

*Index Terms*— HTS dynamo pump, coated conductor, ferromagnetic substrate, thin shell model, Chebyshev spectral method

## I. INTRODUCTION

Electromagnetic induction and nonlinear resistivity of type-II superconductors enable HTS magnetic flux pumps to inject a high DC current into a closed-loop superconducting coil of a magnet and also to continuously compensate the decay of this current [1]. Wireless magnet excitation eliminates excessive cryogenic losses and this is the reason of much recent interest in HTS pumps, see the reviews [2-5]. Dynamo-type pumps, first proposed in [6], have a simple structure: one or several permanent magnets are mounted on a rotating disk and, passing close to a superconducting tape (the stator) induce there a traveling magnetic field wave generating an output voltage with a nonzero average. HTS dynamos have been intensively investigated experimentally; numerical simulations (see the comprehensive recent review [7]) helped to understand the physical mechanism of voltage generation and the impact on the dynamo pump efficiency of various geometrical factors and field-dependent current-voltage relation for the superconductor.

Although losses in coated conductors with magnetic substrates were studied experimentally and numerically in a number of works (see [8-10] and the references therein), our interest is the pump-generated DC voltage. To the best of our knowledge, this issue has not been investigated yet. Here we investigate the influence of a magnetic substrate of the coated conductor employed as a dynamo pump stator on the pump characteristics. Incorporating the recently developed thin shell model of a coated conductor with a ferromagnetic substrate [9] into the pump model ([11], see also [12]), we obtain a system of one-dimensional integro-differential equations. We solve this system using the fast and accurate spectral numerical method [9] and show first that ferromagnetic substrates can increase the generated open-circuit DC voltage. Then, assuming a simplified lumped model of a coil, we simulate its charging during several thousands of rotor revolutions and demonstrate that such substrates accelerate this process. Finally, we show that this time-consuming simulation can be replaced by a simple analytical description of the charging process based on solutions to a few given transport current problems.

Our aim is qualitative characterization of the magnetic substrate impact and, in our model, we employ the simplest constitutive relations: a power current-voltage relation with a constant critical current density for the superconductor and a constant magnetic susceptibility for the substrate. We also neglect currents in all layers of the coated conductor except the superconducting one. Numerical simulations were performed in Matlab R2020b on a PC with the Intel® Core™ i7-9700 CPU 3.00 GHz.

## II. THE MODEL

The HTS dynamo model [1, 11, 12] is developed for a rotating long permanent magnet passing close to a stationary long thin coated conductor strip (fig. 1).

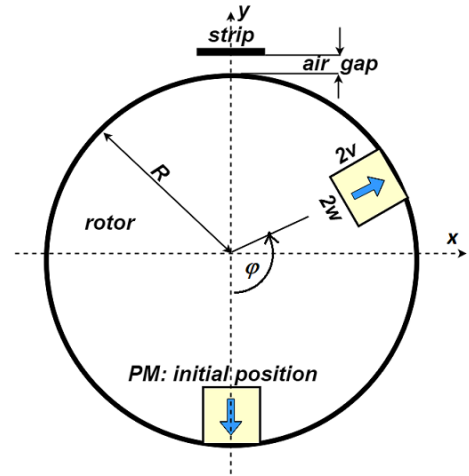

**Fig. 1.** A scheme of an HTS dynamo: the geometry of the problem.

TABLE I. HTS dynamo benchmark parameters.

| | | |
|---|---|---|
| permanent magnet | width, $2w$ | 6 mm |
| | height, $2v$ | 12 mm |
| | active length, $l$ | 12.7 mm |
| | Remanent flux density, $B_r$ | 1.25 T |
| HTS layer | width, $2a$ | 12 mm |
| | thickness | 1 μm |
| | critical current, $I_c$ | 283 A |
| | power value, $n$ | 20 |
| rotor external radius, $R$ | | 35 mm |
| air gap, $d$ | | 3.7 mm |
| frequency of rotation, $f$ | | 4.25 Hz |

In this work we assume the same dynamo configuration and parameters as in [11, 12] (see table I) with one exception: now the coated conductor has a ferromagnetic substrate with magnetic susceptibility $\chi$ and thickness $\delta$; its width $2a$ is equal to that of the HTS layer. We consider two possible orientations of the stator strip: the substrate is further from the rotor than the superconducting layer and vice versa.

Following [9], we consider only two layers of the coated conductor: the thin HTS layer and its substrate. Electrical current is allowed only in the HTS layer, for which we use the infinitely thin approximation, assume the sheet current density is directed along the strip and is the same in all strip cross-sections. The power current-voltage relation, characterizing the HTS layer, is

$$e = e_0 \left| \frac{j}{j_c} \right|^{n-1} \frac{j}{j_c}.$$

Here $j(t,x)$ is the parallel to the $z$-axis sheet current density, $e(t,x)$ is the electric field, $e_0 = 10^{-4}$ V/m, and $j_c = I_c / 2a$ is the field-independent sheet critical current density.

If the substrate material is ferromagnetic, e. g. a Ni-W alloy with the magnetic susceptibility $\chi$ of the order of tens and more, its magnetization influences the current distribution in the superconducting layer and should be taken into account. The thickness of a substrate layer, $30-100$ μm, is small comparing to its width, 4-12 mm. Hence, we can use the thin shell magnetization theory developed by Krasnov [13-15] in terms of "surface magnetization", $\sigma(t,x)$, which is attributed to the mid-surface of the substrate and can be regarded as scalar in the case of a long strip in a perpendicular field. As in [9], we use scaled dimensionless variables

$$\tilde{j} = \frac{j}{j_c}, \quad \tilde{\sigma} = \frac{\sigma}{aj_c}, \quad \tilde{h} = \frac{h}{j_c}, \quad \tilde{e} = \frac{e}{e_0},$$

$$(\tilde{x}, \tilde{y}) = \frac{(x,y)}{a}, \quad \tilde{t} = \frac{t}{t_0}, \quad \tilde{I} = \frac{I}{I_c}, \quad \tilde{V} = \frac{V}{ae_0},$$

where $t_0 = a\mu_0 j_c / e_0$, $\mu_0$ is the magnetic permeability of vacuum, $I$ is the transport current, $V$ is the voltage. Omitting the sign "~" to simplify our notations, we present the coated conductor model in dimensionless form (see [9]),

$$\begin{cases} \kappa^{-1}\sigma(t,x) + \partial_x \left( \frac{1}{2\pi} \int_{-1}^{1} \frac{\sigma(t,x')}{x-x'} dx' \right) + \frac{s}{2} j(t,x) \\ \quad = h_x^e, \\ \partial_t \left( h_y^e + \frac{1}{2\pi} \int_{-1}^{1} \frac{j(t,x')}{x-x'} dx' + \frac{s}{2} \frac{\partial \sigma(t,x)}{\partial x} \right) \\ \quad = \partial_x e(t,x), \\ \int_{-1}^{1} j(t,x) dx = 2I(t), \quad e = |j|^{n-1} j, \end{cases} \quad (1)$$

where $\kappa = \chi\delta/a$ is the dimensionless parameter characterizing the ferromagnetic substrate, $s = -1$ if the substrate is between the rotor and HTS layer and $s = +1$ if the HTS layer is between the rotor and substrate. The field $\mathbf{h}^e = (h_x^e(t,x), h_y^e(t,x))$ is the external magnetic field in the thin strip points. In our case it is the field induced by the rotating permanent magnet. The analytical expressions [16] were used to find this field at each moment in time; see [12] for details of this computation.

The instantaneous voltage on the load contacts can be divided into two parts. First, the voltage generated in the stator and estimated as $V_r(t) = -l\langle e \rangle$. Here $\langle e \rangle = 0.5 \int_{-1}^{1} e(t,x) dx$ is the width-average electric field, and $l$ is the normalized "active length", usually taken equal to the length (in the $z$-axis direction) of the permanent magnet. Second, the voltage generated by the time derivative of the magnetic flux $\Phi$, which depends on geometrical configuration of the electrical circuit and is more difficult to estimate. We note, however, that the cycle-average value of this part is zero for the open-circuit case. For load charging, since the transport current change during one rotor rotation is very small, the cycle-averaged value of $d\Phi/dt$ can usually be neglected too. Hence, as in [11, 12], see also [17], the most important characteristic of an HTS dynamo, the DC voltage, can be calculated as

$$\langle V \rangle = f \int_{t_0}^{t_0+1/f} V_r(t) dt. \quad (2)$$

The one-dimensional integro-differential model of an HTS dynamo, (1)-(2), is simplified in many aspects. The main simplification is the long permanent magnet assumption, which makes possible to consider only a cross-section of the stator but needs the notion of an effective (active) length. This assumption, often employed in dynamo models with non-magnetic substrates (e.g., [1, 11, 12]), enables very fast numerical simulations but provides no realistic description of the closed current loops in the HTS layer, as do more accurate full-dimensional models [18-20]. Comparison of numerical solutions, obtained using the two types of models (see [19], fig. 7) showed, however, that the simplified model still produces a good approximation to the open circuit voltage curve if the strip width does not exceed the permanent magnet length. The thin shell magnetization model we employed for the substrate also makes numerical solution much easier. This simplification is, however, well justified [13-15].

For numerical solution of equations (1) we employed the method presented in [9] and based on the method of lines for integration in time and Chebyshev spectral approximation for discretization in space. The values of unknowns $j$ and $\sigma$ are found in $N+1$ nodes of the Chebyshev mesh $x_k = -\cos(\pi k / N),\ k = 0,...,N.$ This is an extension of the spectral method [12], developed there for HTS dynamos with a non-magnetic stator substrate and shown to be more efficient than all numerical methods considered in [11].

## III. OPEN CIRCUIT VOLTAGE

To model the open circuit conditions, we set $I(t) = 0$ and compare voltages computed for different ferromagnetic substrates characterized by their values of parameter $\kappa$. To avoid the transient effects, we present our simulation results for the second rotor rotation cycle. The computations were performed using the Chebyshev mesh with $N = 200$ and the computation time was about 20 seconds per cycle. For $N = 100$ this time was less than 5 seconds and the difference in computed DC voltages was less than 1%.

First, let the HTS layer be on the rotor side of the strip, so in the model (1) we set $s = 1$. For small $\kappa$ the influence of the magnetic substrate is negligible, the voltage curve and the DC voltage are similar to those for a non-magnetic substrate. As this parameter increases, the voltage curve changes and the two negative peaks become deeper (fig. 2). The DC voltage changes from -10.1 to -16 $\mu V$ (fig. 3, top) and saturates: the results for $\kappa = 10$ and $\kappa = 100$ are close. We conclude that, if the HTS layer is between the rotor and the substrate, the magnetic substrate can significantly increase the DC voltage (its absolute value, the sign is unimportant).

In case of the opposite orientation (the magnetic substrate is between the rotor and HTS layer, $s = -1$) the substrate shields the magnetic field and the DC voltage decreases; it becomes negligible for large $\kappa$ (fig. 3, bottom). A similar influence of the ferromagnetic slice inserted before or behind the coated conductor dynamo stator has been observed experimentally and simulated numerically in [21].

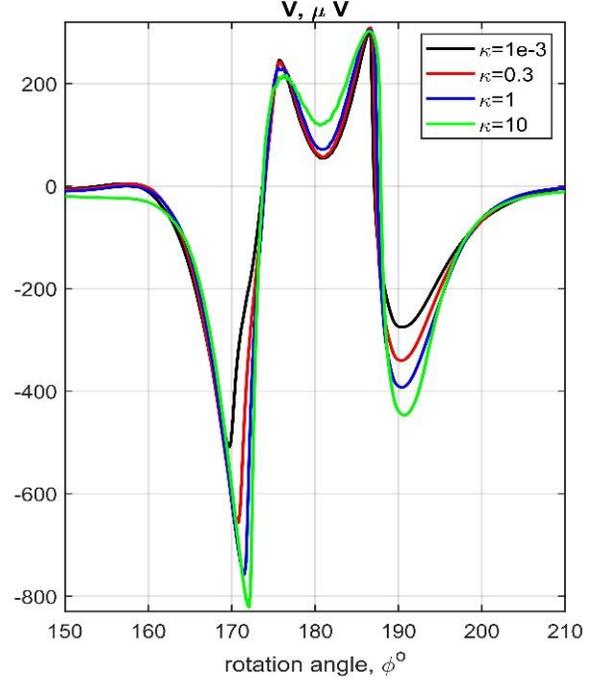

**Fig. 2.** The voltage $V_r$ during the second rotation cycle: the influence of ferromagnetic substrate.

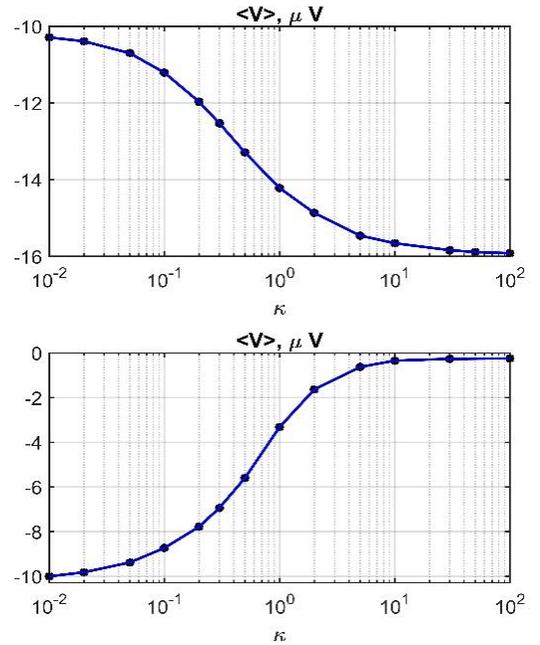

**Fig. 3.** Influence of magnetic substrate on the generated DC voltage $<V>$. Top: HTS layer is between the rotor and the substrate. Bottom: the substrate is between the rotor and HTS layer.

## IV. CHARGING A SUPERCONDUCTING COIL

We now assume the superconducting layer is between the rotor and substrate (the DC voltage is increased due to the ferromagnetic substrate) and consider first the problem with a given transport current. Let the current be changed linearly from zero to a prescribed value $I$ during the first cycle and then remain constant. The DC voltage was now computed for the third cycle. Our numerical simulations showed that for each $\kappa$ the voltage $<V>$ depends almost linearly on the transport current $I$ (fig. 4).

Denoting by $V_0(\kappa)$ the open-circuit DC voltage, we approximate this dependance as

$$<V>=V_0(\kappa)-R_{\text{eff}}(\kappa)I, \qquad (3)$$

where $R_{\text{eff}}$ is a constant playing the role of an effective stator resistance. The voltage becomes zero at almost the same transport current, $I_0(\kappa)$, for all $\kappa$. We find $R_{\text{eff}}(\kappa)=V_0(\kappa)/I_0(\kappa)$ (table II).

To simulate charging a superconducting magnet by an HTS dynamo we consider, as in [22] for the case of a nonmagnetic substrate, the dynamo in a closed circuit with a simplified lumped load having the resistivity $R$ and inductivity $L$. We use the same values of these parameters as in [22]: $R=0.88\,\mu\Omega$, $L=0.24\,\text{mH}$.

**Table II.** Dependance of the main pump characteristics on the ferromagnetic substrate.

| $\kappa$ | 0.01 | 0.3 | 1 | 10 |
|---|---|---|---|---|
| $V_0$ ($\mu$V) | -10.3 | -12.5 | -14.2 | -15.7 |
| $I_0$ (A) | -31.1 | -34.0 | -34.0 | -31.1 |
| $R_{\text{eff}}$ ($\mu\Omega$) | 0.330 | 0.369 | 0.418 | 0.503 |
| $I_{\text{sat}}$ (A) | 8.5 | 10 | 10.9 | 11.4 |
| $\tau$ (s) | 198 | 192 | 185 | 174 |

The circuit equation

$$V_0(\kappa)-R_{\text{eff}}(\kappa)I = RI + LdI/dt$$

yields

$$I = I_{\text{sat}}\left[1-\exp(-t/\tau)\right], \qquad (4)$$

where the saturation current and the characteristic charging time are, respectively,

$$I_{\text{sat}} = V_0/(R_{\text{eff}}+R), \quad \tau = L/(R_{\text{eff}}+R).$$

These parameters depend on $\kappa$, see table II.

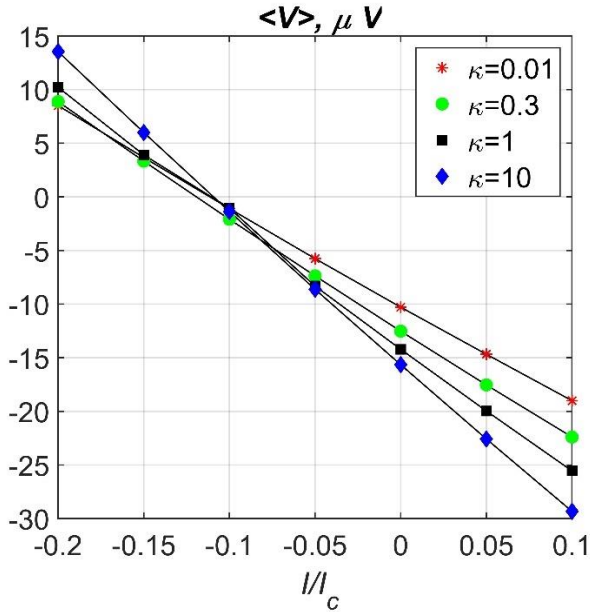

**Fig. 4.** Pump-generated DC voltage as a function of transport current for different ferromagnetic substrates.

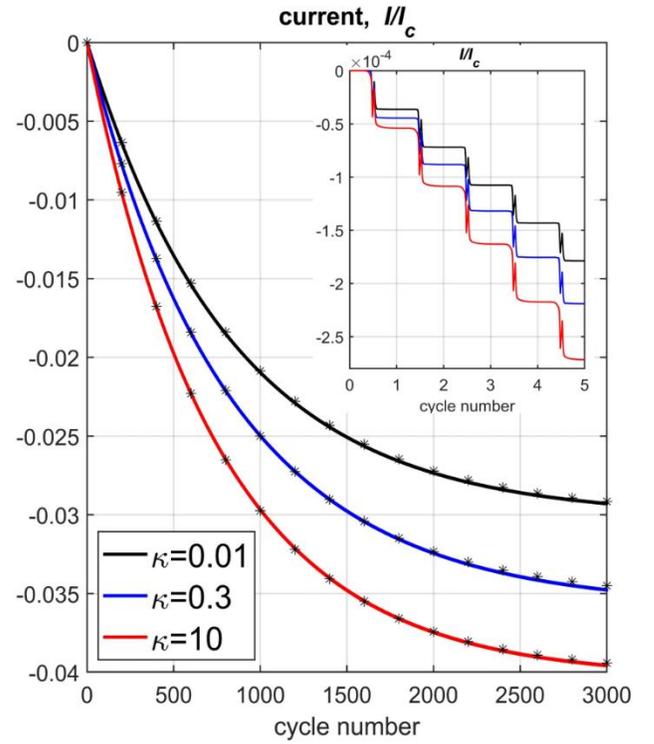

**Fig. 5.** Numerical simulation of load charging by dynamos with different stator substrates. Solid lines – solutions to the model (1),(5). Black stars indicate the corresponding analytical solutions (4). The inset illustrates the current ripples, accompanying the process but not seen in the current curves for 3000 cycles.

Solution (4) completely ignores oscillations of the pump voltage during each cycle. On the other hand, not using the linear approximation (3), we can supplement our model (1) by the differential equation

$$RI + L\, dI/dt = V_r(t), \qquad (5)$$

in which the right hand side takes into account only the part of the pump voltage ripples related to the resistance of the stator but, anyway, has the cycle-averaged value equal to that of the total voltage. It is easy to incorporate the evolutionary equation (5) into our numerical scheme that uses the method of lines for integration in time. The employed numerical method is efficient and we were able to model several thousands of cycles and compare the analytical solution (4) with the numerical solution of the system (1),(5); see fig. 5. The two solutions practically coincide, which suggests that in modeling charging current the ripples can be ignored. The presence of ripples can, however, increase the AC loss.

## VI. Conclusion

The benchmark HTS dynamo pump problem [11] was in this work extended to the case of the pump stator made of a coated conductor with a ferromagnetic substrate. Such a substrate changes the superconducting current density distribution. To our knowledge, the impact of a magnetic substrate on the dynamo pump performance has not been studied yet.

In our work [9], using the thin shell magnetization theory [13-15], we presented a new model of a coated conductor with a ferromagnetic substrate and an efficient Chebyshev spectral numerical method. The model makes use of the high width-to-thickness ratio of the substrate and superconducting layers and is much simpler than the previously proposed two-dimensional models (for which the high aspect ratio presents a difficulty). In this work we applied such approach to modeling the HTS dynamo pumps and showed that magnetic stator substrate can significantly increase the pump generated voltage and accelerate contactless charging of a coil if the superconducting layer is oriented towards the rotated permanent magnet. For the opposite stator orientation, the magnetic field is shielded by the substrate and the pump generated voltage decreases.

For a given transport current, determining the pump generated voltage using our model and numerical scheme takes less than a minute on a PC. Results of such simulations can be used to determine also the effective lumped model parameters of the pump and replace simulation of charging a coil during thousands rotor rotations by a simple analytical formula.

## References

[1] R. Mataira, M. Ainslie, R. Badcock, and C. Bumby, "Origin of the DC output voltage from a high-Tc superconducting dynamo," *Appl. Phys. Lett.,* vol. 114, no. 16, 2019, Art. no. 162601.

[2] Y. Zhai, Z. Tan, X. Liu, B. Shen, T. A. Coombs, and F. Wang, "Research progress of contactless magnetization technology: HTS flux pumps," *IEEE Trans. Appl. Supercond.,* vol. 30, no. 4, 2020, Art. no. 4602905.

[3] T. A. Coombs, J. Geng, L. Fu, and K. Matsuda, "An overview of flux pumps for HTS coils," *IEEE Trans. Appl. Supercond.,* vol. 27, no. 4, 2017, Art. no. 4600806.

[4] Z. Wen, H. Zhang, and M. Mueller, "High temperature superconducting flux pumps for contactless energization," *Crystals,* vol. 12, no. 6, 2022, Art. no. 766.

[5] W. Wang, J. Wei, C. Yang, C. Wu, and H. Li, "Review of high temperature superconducting flux pumps," *Superconductivity,* vol. 3, 2022, Art.no. 100022.

[6] C. Hoffmann, D. Pooke, and A. D. Caplin, "Flux pump for HTS magnets," *IEEE Trans. Appl. Supercond.,* vol. 21, no. 3, 2010, pp. 1628-1631.

[7] M. D. Ainslie, "Numerical modelling of high-temperature superconducting dynamos: A review," *Superconductivity,* vol. 5, Mar. 2023, Art. no. 100033.

[8] Y. Mawatari, "Magnetic field distributions around superconducting strips on ferromagnetic substrates," *Phys. Rev. B,* vol. 77, no. 10, Art. no. 104505, 2008.

[9] L. Prigozhin and V. Sokolovsky, "Thin shell model of a coated conductor with a ferromagnetic substrate," *IEEE Tran. Appl. Supercond.,* vol. 33, no. 4, 2023, Art.no. 6601310.

[10] Y. Statra, H. Menana, and B. Douine, "Integral modeling of AC losses in HTS tapes with magnetic substrates," *IEEE Tran. Appl. Supercond.,* vol. 32, no. 2, 2021, Art. no. 5900407.

[11] M. Ainslie *et al.*, "A new benchmark problem for electromagnetic modelling of superconductors: the high-$T_c$ superconducting dynamo," *Supercond. Sci. Technol.,* vol. 33, no. 10, 2020, Art. no. 105009.

[12] L. Prigozhin and V. Sokolovsky, "Fast solution of the superconducting dynamo benchmark problem," *Supercond. Sci. and Technol.,* vol. 34, no. 6, 2021, Art. no. 065006.

[13] I. P. Krasnov, "Integral equation for magnetostatic problems with thin plates or shells," *Sov. Phys. Tech. Phys.,* vol. 22, no. 7, 1977, pp. 811-817.

[14] I. P. Krasnov, "Solution of the magnetostatic equations for thin plates and shells in the slab and axisymmetric cases," *Sov. Phys. Tech. Phys.,* vol. 27, no. 5, 1982, pp. 535-538.

[15] I. P. Krasnov, *Computational Methods of Ship Magnetism and Electric Engineering*. Leningrad, Russia: Sudostroenie [in Russian], 1986.

[16]  E. P. Furlani, *Permanent Magnet and Electromechanical devices: Materials, Analysis, and Applications*. New York: Academic press, 2001.

[17]  R. Mataira-Cole, "*The Physics of the High-Temperature Superconducting Dynamo and No-Insulation Coils*" (Doctoral dissertation, Open Access Te Herenga Waka-Victoria University of Wellington) 2022,

[18]  A. Ghabeli, E. Pardo, and M. Kapolka, "3D modeling of a superconducting dynamo-type flux pump," *Scientific Reports,* vol. 11, no. 1, 2021, Art. no. 10296.

[19]  L. Prigozhin and V. Sokolovsky, "Two-dimensional model of a high-$T_c$ superconducting dynamo," *IEEE Trans. Appl. Supercond.,* vol. 31, no. 3, 2021, Art. no. 065006.

[20]  V. Sokolovsky and L. Prigozhin, "Hermite-Chebyshev pseudospectral method for inhomogeneous superconducting strip problems and magnetic flux pump modeling," *Supercond. Sci. Technol.,* vol. 35, no. 2, 2021, Art. no. 024002.

[21]  J. Li *et al.*, "Influence of ferromagnetic slice on the charging performance of a through-wall HTS flux pump employing a magnetic coupler," *Supercond. Sci. Technol.,* vol. 35, no. 7, 2022, Art. no. 075008.

[22]  A. Ghabeli, M. Ainslie, E. Pardo, L. Quéval, and R. Mataira, "Modeling the charging process of a coil by an HTS dynamo-type flux pump," *Supercond. Sci. Technol.,* vol. 34, no. 8, 2021, Art. no. 084002.